\documentclass[aps,prl,twocolumn,superscriptaddress,hyperref]{revtex4}

\usepackage{graphicx}
\usepackage{latexsym}
\usepackage{amsmath}
\usepackage{amssymb}

\newcommand*{\bk}{\mathbf{k}}
\newcommand*{\bq}{\mathbf{q}}

\newcommand{\vect}[1]{{\mathbf #1}}

\begin{document}


\title{Finite temperature phase diagram of a polarised Fermi condensate}

\author{M. M. Parish}
\affiliation{Cavendish Laboratory, JJ Thomson Avenue, Cambridge,
CB3 0HE, United Kingdom}

\author{F. M. Marchetti}
\affiliation{Cavendish Laboratory, JJ Thomson Avenue, Cambridge,
CB3 0HE, United Kingdom}

\author{A. Lamacraft}
\affiliation{Rudolf Peierls Centre for Theoretical Physics, 1 Keble
Road, Oxford OX1 3NP, UK and All Souls College, Oxford.}

\author{B. D. Simons}
\affiliation{Cavendish Laboratory, JJ Thomson Avenue, Cambridge,
CB3 0HE, United Kingdom}

\date{\today}

\begin{abstract}
The two-component Fermi gas is the simplest fermion system
displaying superfluidity, and as such is relevant to topics ranging
from superconductivity to QCD. Ultracold atomic gases provide an
exceptionally clean realisation of this system, where interatomic
interactions and atom spin populations are both independently
tuneable.
Here we show that the finite temperature phase diagram contains a
region of phase separation between the superfluid and normal states
that touches the boundary of second-order superfluid transitions at
a tricritical point, reminiscent of the phase diagram of
$^3$He-$^4$He mixtures. A variation of interaction strength then
results in a line of tricritical points that terminates at zero
temperature on the molecular Bose-Einstein condensate (BEC) side. On
this basis, we argue that tricritical points are fundamental to
understanding experiments on polarised atomic Fermi gases.
\end{abstract}

\maketitle



Over the past decade, experimental progress in the field of cold
atomic gases has resulted in unprecedented control over pairing
phenomena in two-component Fermi gases. The ability to vary the
effective interaction between atoms using magnetically tuned
Feshbach resonances has already permitted the experimental
investigation of the crossover from a BEC of diatomic molecules to
the Bardeen-Cooper-Schrieffer (BCS) limit of weakly-bound Cooper
pairs of fermionic
atoms~\cite{regal2004,zwierlein2004,chin2004,bourdel2004,kinast2004,zwierlein2005}.
A natural extension of these studies is an exploration of the Fermi
gas with imbalanced spin populations, especially since this system
has a far richer phase diagram than the equal spin case.  As well as
exhibiting a quantum phase transition between the superfluid and
normal states, the polarized Fermi gas has been predicted to possess
exotic superfluid phases such as the inhomogeneous
Fulde-Ferrell-Larkin-Ovchinnikov (FFLO)
state~\cite{Fulde1964,larkin1965}, where the pairing of fermions
occurs at finite centre-of-mass momentum, and the deformed Fermi
surface state~\cite{sedrakian2005}. The exact nature of the
superfluid states for the polarised Fermi gas is still the subject
of considerable debate. However, atomic gases provide an ideal
testing ground for this system, since the particle numbers can be
varied independently from all other experimental parameters, and
pioneering experiments have recently been
performed~\cite{zwierlein2006,partridge2006,zwierlein2006_2,shin2006}.
Contrast atomic gases with the case of superconductors, where the
magnetic field used to generate a spin imbalance (via the Zeeman
effect) also couples to orbital degrees of freedom.

In this work, we elucidate the finite temperature phase diagram of a
polarised Fermi gas. While much insight has been gained from
previous theoretical
studies~\cite{bedaque2003,carlson2005,pao2005,son2005,Mizushima2005,sheehy2006,%
mannarelli2006,pieri2005,liu2006,hu2006,chien2006,gu2006,martikainen2006,%
iskin2006,desilva2006,haque2006,yi2006,kinnunen2006}, so far a key
ingredient of the phase diagram has been overlooked: the tricritical
point, at which the phase transition between superfluid and normal
states switches from first to second order. By determining the
behaviour of the tricritical point as a function of interaction
strength, we can completely characterise the topology of the phase
diagram without recourse to an extensive numerical treatment.
Specifically, we shall focus on the uniform, infinite system, and
concern ourselves almost exclusively with the phase boundary between
the normal and homogeneous superfluid states. We will, however,
discuss the ramifications of the inferred phase diagram for the
trapped system.

\section{Formalism}
Experiments to date exploit wide Feshbach resonances and are thus
well described by the simplest single-channel Hamiltonian, where the
two fermion species interact via an attractive contact potential
\begin{multline}\label{eq:ham}
  \hat{H} - \mu_{\uparrow} \hat{n}_{\uparrow} - \mu_{\downarrow}
  \hat{n}_{\downarrow} = \sum_{\vect{k}} \sum_{\sigma =
  \uparrow,\downarrow} \left(\epsilon_{\vect{k}} - \mu_{\sigma}\right)
  c_{\vect{k} \sigma}^\dag c_{\vect{k} \sigma} \\
   + \frac{g}{V} \sum_{\vect{k},\vect{k}',\vect{q}}
  c_{\vect{k}+\vect{q}/2 \uparrow}^\dag c_{-\vect{k}+\vect{q}/2
  \downarrow}^\dag c_{-\vect{k}'+\vect{q}/2 \downarrow}
  c_{\vect{k}'+\vect{q}/2 \uparrow}\; .
\end{multline}
Here, $\epsilon_{\vect{k}}={\vect{k}}^2/2m_{f}$ (we set $\hbar=1$ and $k_B=1$),
$V$ is the volume, and we define the chemical potential $\mu$ and
`Zeeman' field $h$ such that $\mu_{\uparrow} = \mu + h$ and
$\mu_{\downarrow} = \mu - h$. At present, only pairing between
different hyperfine species of the \emph{same} atom has been
explored experimentally, so we restrict ourselves to a single mass
$m_f$. The interaction strength $g$ is expressed in terms of
the s-wave scattering length $a$ using the prescription:
\begin{equation*}
  \frac{m_f}{4\pi a} = \frac{1}{g} + \frac{1}{V}\sum_{\vect{k}}
  \frac{1}{2\epsilon_{\vect{k}}} \; .
\end{equation*}
We also derive the Fermi momentum using the average density $n/2
\equiv (n_{\uparrow}+n_{\downarrow})/2$, so that $k_F = (3\pi^2
n)^{1/3}$.
Throughout our calculations, we will keep $n$ fixed.

The full phase diagram is parameterised by just a few observables:
the temperature $T\equiv 1/\beta$, the interaction strength
$1/k_Fa$, and the density difference or `magnetisation' $m\equiv
n_{\uparrow}-n_{\downarrow}$. To determine the position of the phase
boundaries, we must minimise the mean-field free energy density
\begin{multline}\label{eq:energy}
  \Omega^{0} = -\frac{\Delta^2}{g} +
  \frac{1}{V}\sum_{\vect{k}} (\xi_{\vect{k}} - E_{\vect{k}}) \\
  -\frac{1}{\beta V}\sum_{\vect{k}}
  \left[\ln\left(1+e^{-\beta(E_{\vect{k}}-h)}\right)
   + \ln\left(1+e^{-\beta(E_{\vect{k}}+h)}\right) \right],
\end{multline}
with respect to the BCS order parameter $\Delta$, where
$\xi_{\vect{k}}=\epsilon_{\vect{k}}-\mu$ and
$E_{\vect{k}}=\sqrt{\xi_{\vect{k}}^2+\Delta^2}$.  Such a mean-field
analysis provides a reasonable description of the zero temperature
phase diagram, but at finite temperature, it neglects the
contribution of non-condensed pairs to both the density $n =
-\partial \Omega/\partial \mu$ and magnetisation $m = -\partial
\Omega/\partial h$. This contribution is necessary to approach the
transition temperature of an ideal Bose gas in the molecular limit,
and can be included in the non-condensed phase ($\Delta=0$) through
the Nozi\`eres-Schmitt-Rink (NSR) fluctuation correction to the
energy~\cite{nozieres1985}
\begin{equation} \label{fluct}
\left.\Omega^{1}\right|_{\Delta = 0} = \frac{1}{\beta V}
\sum_{\vect{q},i\omega}\ln\Gamma^{-1}(\vect{q},i\omega) \; ,
\end{equation}
with
\begin{multline}
\Gamma^{-1}(\vect{q},i\omega) = -\frac{1}{g} \\
- \frac{1}{2V}\sum_{\vect{p}}
\frac{\tanh{[\frac{\beta}{2}(\xi_{\vect{p}} + h)] +
\tanh[\frac{\beta}{2}(\xi_{\vect{p}+\vect{q}} - h)]}}{i\omega +
\xi_{\vect{p}} + \xi_{\vect{p}+\vect{q}}} \; .
\end{multline}
This gives an estimate of the effect of pair fluctuations on
the second order phase boundary (but not the first order boundary, where
$\Delta \neq 0$).
\begin{figure}
\centering
\includegraphics[width=0.45\textwidth]{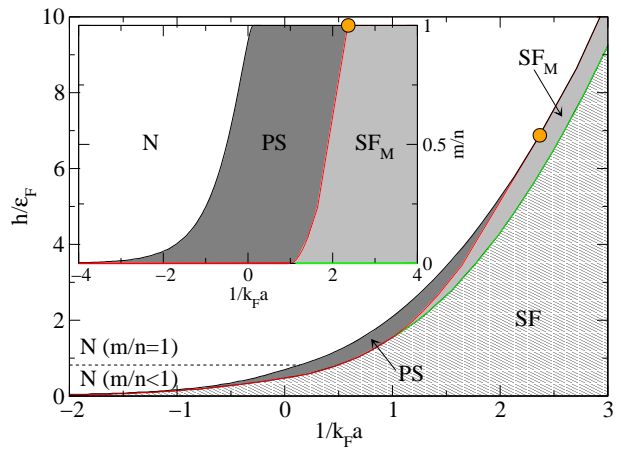}
\caption{The zero temperature phase diagram within mean-field theory
for both Zeeman field $h/\varepsilon_F$ and magnetisation $m/n$
(inset) versus interaction $1/k_Fa$. There are four different
phases: the normal (N) state, the phase-separated (PS) state, the
ordinary superfluid (SF) and the magnetised superfluid
(SF$_{\textrm{M}}$). Above the line $h/\varepsilon_F = 2^{-1/3}$,
the normal state is completely polarised ($m/n=1$). The red and
black lines enclosing the PS state are both first-order phase
boundaries, while the SF$_{\textrm{M}}$-N transition is
second-order, and the SF-SF$_{\textrm{M}}$ transition (green line)
is at least third-order. The tricritical point is represented by
orange circles at $1/k_Fa=2.368$ with $h/\varepsilon_F = 6.876$ or
$m/n=1$. \label{fig:zeroT}}
\end{figure}

\section{Phase diagram for the uniform case}
Considerable insight can be gained by first examining the zero
temperature mean-field phase diagram, as shown in
Fig.~\ref{fig:zeroT}.
The general structure parallels that of the two-channel case found in Ref.~\cite{sheehy2006}.
Since there is a gap in the quasiparticle
excitation spectrum $E_{\vect{k}}$ of the unpolarised superfluid,
the superfluid ground state will remain unchanged for
$h<\mathrm{min}_{\bk} E_{\bk}$. We see that the $m=0$ superfluid
line in the inset of Fig.~\ref{fig:zeroT} corresponds to an
\emph{area} in the $h/\varepsilon_F$ versus $1/k_Fa$ diagram, which
expands as $1/k_Fa$ increases. A key feature of the strong coupling
side is that for $1/k_F a\gtrsim 1$ the superfluid state is able to
sustain a finite population of majority quasiparticles. This
``gapless''~\cite{pao2005,son2005} superfluid phase is only stable
for $\mu<0$ and it thus possesses only one Fermi surface. In the
extreme BEC limit, this state is straightforwardly understood as an
almost ideal mixture of bosonic pairs and fermionic quasiparticles.
However, as we move towards unitarity, the bosons and fermions begin
to interact more strongly, leading eventually to a first-order phase
transition to the normal state. Here, a system with fixed $m$ will
undergo phase separation into normal and superfluid regions if
$m_N<m<m_S$, where $m_{N,S}$ denotes the magnetisation in the normal
and superfluid phases at $h_c$, the critical field for the
first-order transition. In the BCS limit ($\mu = \varepsilon_F$),
$h_c=\Delta/\sqrt{2}$ which is less than the quasiparticle gap, so
the superfluid state is unmagnetised $m_S=0$, and phase separation
occurs for arbitrarily low magnetisation, consistent with
Ref.~\cite{bedaque2003}. For the moment we neglect the FFLO state, but will
return to this point later.


A crucial observation is that the line $m/n=1$ to the right of the
region of phase separation can be thought of as a continuous zero
temperature transition at which the condensate is totally depleted.
It is thus natural to identify the point on $m/n=1$ where phase
separation starts as a tricritical point. Indeed a Landau
expansion of the free energy both confirms this and identifies
the tricritical point at $1/k_Fa=2.368$.
\begin{figure}
\centering
\includegraphics[width=0.45\textwidth]{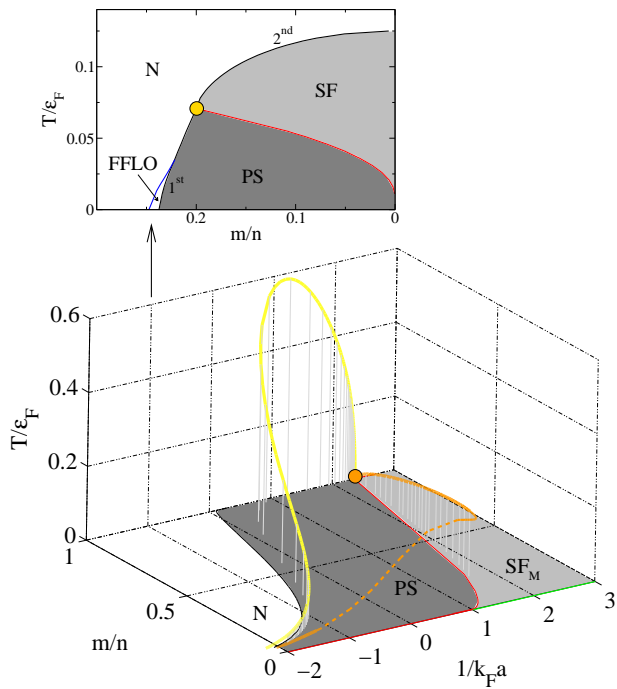}
\caption{Finite temperature phase diagram as a function of
magnetisation $m/n$ and interaction $1/k_Fa$. The plane at
temperature $T=0$ is the phase diagram in Fig.~\ref{fig:zeroT}. The
yellow line represents the locus of tricritical points calculated in
the mean-field approximation, while the orange tricritical line
corresponds to mean-field theory plus pair fluctuations.  The
fluctuation correction breaks down in the unitarity regime $-1 <
1/k_Fa < 1$, and is thus shown as a dotted line.  The slice at
$1/k_Fa = -1$ is based on a mean-field calculation and it shows the
region of phase separation terminating in a tricritical point
(yellow circle) at finite temperature, followed by a second-order
phase transition from the superfluid to normal state. Note that the
boundary between the FFLO and normal states (blue line) defines
a small region of FFLO phase confined to the BCS side of the crossover, as explained in the text.}
\label{fig:tricritical}
\end{figure}
\begin{figure}
\centering
\includegraphics[width=0.45\textwidth]{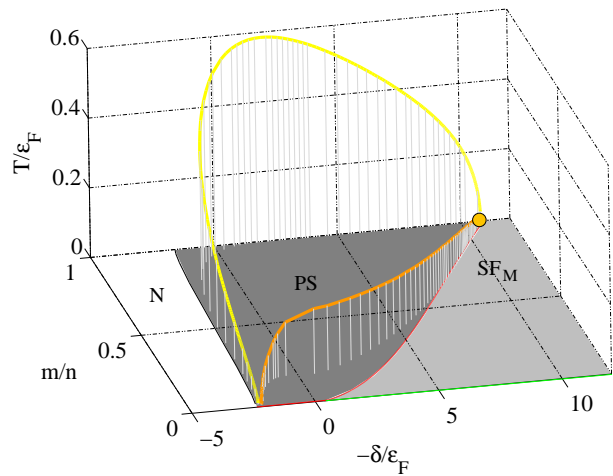}
\caption{Finite temperature phase diagram for the two-channel model
of a narrow Feshbach resonance, where the coupling between open and
closed channels is weak: $\gamma = 0.1$.  The effective interaction
is parameterised by the detuning $\delta/\varepsilon_F$.
The colour scheme for tricritical lines is the same as in
Fig.~\ref{fig:tricritical}.}
\label{fig:tricritical_2chan}
\end{figure}

With this background, we now turn to the analysis of the fate of the
tricritical point when temperature is finite, beginning with the
mean-field description.  It is well known that there exists a finite
temperature tricritical point in the BCS limit $1/k_Fa\to -\infty$,
which is a natural consequence of having a first-order transition
from the superfluid to normal state at $T=0$ and a second-order
transition at $m=0$. First studied by Sarma in the context of
superconductivity in the presence of a magnetic
field~\cite{sarma1963}, the BCS tricritical point is located at
$(T_{\textrm{crit}}/\Delta,h_{\textrm{crit}}/\Delta) =
(0.3188,0.6061)$~\cite{casalbuoni2004}, where $\Delta =
\frac{8}{e^2}\varepsilon_F\exp\left[-\pi/2|k_F a|\right]$ (i.e. at
weak coupling all energies scale with $\Delta$). This corresponds to
a magnetisation $m=2\nu(\varepsilon_F)h_{\textrm{crit}}$, where
$\nu(\varepsilon_F) = m_f^{3/2} \sqrt{\varepsilon_F}/ \sqrt{2}\pi^2$
is the Fermi surface density of states.  To investigate how the BCS
tricritical point is related to the one at zero temperature, we must
develop a perturbative expansion of Eq.~(\ref{eq:energy}) for small
$\Delta$ and general $1/k_Fa$. Doing so, one finds
(Fig.~\ref{fig:tricritical}) that the tricritical point at $m/n=1$
is connected to that in the BCS limit by a line of tricritical
points that passes through a maximum somewhere in the `unitarity'
regime $-1<1/k_Fa<1$. Moreover, for any given value of $1/k_Fa\leq
2.368$, the $(T/\varepsilon_F,m/n)$ phase diagram is highly
reminiscent of  the $^3$He-$^4$He system, with $m/n$ playing the
role of the fraction of $^3$He. This is not surprising, as the
finite $m$ system corresponds in general to a mixture of bosonic
pairs and fermionic quasiparticles. Note that even the gapped
superfluid can be magnetised at finite temperature due to thermal
excitation of quasiparticles. Of course, at $m=0$ the transition
into the superfluid state is second order at any point in the
BCS-BEC crossover.

It is interesting to examine how the FFLO phase fits in with the
basic topology of the phase diagram. In the BCS limit, we already
know that the point where the FFLO-normal phase boundary meets the
normal-superfluid boundary asymptotes to the tricritical
point~\cite{casalbuoni2004}.
Assuming that the transition from the FFLO state to the normal state
is second-order (although Ref.~\cite{combescot2004} found it to be
weakly first order, this will make a relatively small difference),
and performing a mean-field analysis, we find that the FFLO point of
intersection leaves the finite temperature tricritical point with
increasing interaction (see Fig.~\ref{fig:tricritical}), leading
eventually to the extinction of the FFLO phase at $k_Fa=-0.35$.
Note that although this treatment is somewhat approximate, as we
have taken the SF-FFLO boundary to be the same as the SF-N boundary
in the absence of FFLO, the point of intersection will coincide with
that derived from a complete mean-field analysis. Moreover, despite
all our assumptions, we expect the detachment of the point of
intersection from the tricritical point and the eventual
disappearance of FFLO to be robust features, since in the BEC regime
we essentially have a mixture of bosons and fermions.

The inclusion of the fluctuation contribution Eq.~(\ref{fluct}) is crucial
for recovering the extreme BEC limit, where it is clear that
the (second-order) transition temperature asymptotes to
$T_{\mathrm{BEC}}(m)=T_{\mathrm{BEC}}\left(1-m/n\right)^{2/3}$ (with
$T_{\mathrm{BEC}}\sim 0.218 \varepsilon_F$), the ideal BEC
temperature of a gas of bosons of density $n_\downarrow=(n-m)/2$ and
mass $2m_f$. More importantly, we find that fluctuations shift the
mean-field tricritical line to lower temperatures and magnetisations
on the BEC side, while leaving the tricritical points on the BCS side
largely unchanged, as expected.
However, in a broad region around unitarity, we find that the
approximation underlying Eq.~(\ref{fluct}) generally leads to
non-monotonic behavior of $m(h)$, with $m(h>0)<0$ for small $h$.
We interpret this behaviour as a breakdown of the
NSR treatment, yielding an unphysical compressibility matrix
$-\partial^2\Omega/\partial \mu_{\sigma}\partial \mu_{\sigma'}$ that
is not positive semi-definite.

To address this problem, we note that the NSR scheme is a controlled
approximation when we introduce resonant scattering with a finite width,
with the width being a small fraction of the Fermi energy~\cite{andreev2004}.
The simplest such description is provided by the two-channel
model~\cite{timmermans2001,holland2001}. The two-channel description
of scattering depends upon two parameters: a detuning
$\delta/\varepsilon_F$ describing the distance from the resonance,
and a width $\gamma$ of the resonance measured in units of the Fermi
energy. The one-channel description is recovered in the
$\gamma\to\infty$ limit, while the treatment of Gaussian
fluctuations is essentially perturbative in $\gamma$, with
$\Gamma^{-1}$ in Eq.~(\ref{fluct}) being replaced with
$\frac{\bq^2}{4m}-i\omega_m+\gamma\Gamma^{-1}(\bq,i\omega_{m})$, so
in this case the NSR treatment is expected to be accurate. The
resulting phase diagram is shown in
Figure~\ref{fig:tricritical_2chan}. The zero temperature phase
diagram coincides with the result of Ref.~\cite{sheehy2006}. With
fluctuations accounted for, and for sufficiently small $\gamma$, we now
find a well-behaved line of tricritical points spanning the
crossover region. We expect that the true phase boundary at
$\gamma\to\infty$ is qualitatively similar.

\begin{figure}
\centering
\includegraphics[width=0.42\textwidth]{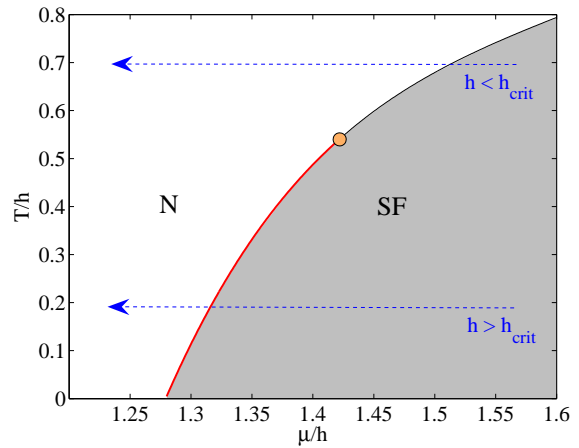}
\caption{Phase diagram at $1/k_Fa=0$ in the $\mu/h$-$T/h$ plane.
The red and black lines are first- and second-order phase
boundaries, respectively.  The arrows at constant $T/h$ represent
the trajectories followed when going from the centre to the
edges of a trapped gas. The two trajectories correspond to two
different magnetisations of the gas: one greater and one less than
the tricritical point $h_{\textrm{crit}}$.} \label{fig:tri_trap}
\end{figure}

\section{Implications for experiment}
We now discuss the consequences of our results for trapped gases
studied in experiment. Modeling the trapped gas by the local density
approximation (LDA), the spatial dependence of the density induced
by the trapping potential $V(\vect{r})$ is accounted for by a
spatially-varying chemical potential $\mu(\vect{r}) = \mu -
V(\vect{r})$, with $h$ kept constant. In the $\mu/h$-$T/h$ plane, we
thus move on a horizontal line (see Fig.~\ref{fig:tri_trap}).  At
sufficiently low temperatures, a trapped gas will consist of a
superfluid core surrounded by the normal state. The transition
between normal and superfluid states in the trap can be either
second or first order, depending on whether $T/h$ is above or below
the tricritical point. Moreover, as long as the temperature is
non-zero, we can always find a sufficiently small $h$ so that $T/h$
lies above the tricritical point.  This leads us to a key point: if
a trapped gas at a given temperature and magnetisation has a
first-order transition between its normal and superfluid phases,
then we will \emph{always} cross the tricritical point by decreasing
the magnetisation at fixed temperature.

We emphasise that there are qualitative differences between first
and second order transitions in a trap: the former yields a
discontinuity in the density and magnetization at the phase
interface, resulting in a form of phase separation as seen in recent
experiments~\cite{zwierlein2006,partridge2006,zwierlein2006_2,shin2006},
while the latter possesses a density that varies smoothly in space.
Therefore, the magnetisation and temperature at which a tricritical
point is crossed should be detectable experimentally. In fact, a
critical magnetisation for the onset of phase separation in a trap
has been observed experimentally~\cite{partridge2006}, and a
calculation by Chevy supports the idea that this coincides with
crossing a tricritical point~\cite{chevy2006}. In addition, the
order of the transition will have an impact on experiments that use
phase separation as a signature of
superfluidity~\cite{zwierlein2006_2}.

The presence of a first-order transition in the trap can be even
more pronounced if the density discontinuities result in a breakdown
of LDA. Experiments on highly elongated traps already provide
evidence for such a breakdown~\cite{partridge2006}, and one requires
the addition of surface energy terms at the phase interface to
successfully model the trapped density
profiles~\cite{desilva2006_2}.

An outstanding issue is the experimental detection of the gapless
SF$_{\textrm{M}}$ phase.
%
%
While optically probing the momentum distribution of the minority
species is one promising method for detecting
SF$_{\textrm{M}}$~\cite{yi2006_3}, another possibility is to study
density correlations using, for example, shot noise experiments as
suggested in Ref.~\cite{altman2004}. A simple mean-field calculation
gives (for the uniform system):
\begin{align*}
 C_{\uparrow\downarrow}(\vect{k_1},\vect{k_2}) & \equiv
 \langle \hat{n}_{\uparrow}(\vect{k_1}) \hat{n}_{\downarrow}(\vect{k_2})\rangle -
 \langle \hat{n}_{\uparrow}(\vect{k_1}) \rangle \langle
 \hat{n}_{\downarrow}(\vect{k_2})\rangle \\
 & = \delta_{\vect{k_1},-\vect{k_2}}
 \frac{\Delta^2}{4E_{\vect{k_1}}^2}
 [1-f(E_{\vect{k_1}}+h)-f(E_{\vect{k_1}}-h)]^2
\end{align*}
where $f(E)$ is the Fermi-Dirac distribution. At $T=0$, the result
is a `hole' in the correlation function for momenta less than the
Fermi wavevector of the majority quasiparticles. Such a measurement
would therefore constitute both a confirmation of the
SF$_{\textrm{M}}$ phase and a vivid demonstration of the blocking
effect of quasiparticles on $\left(+\bk,-\bk\right)$ pairing.

In conclusion, we have determined the structure of the finite
temperature phase diagram of the two component Fermi gas, as a
function of both interaction strength and population imbalance,
finding a region of phase separation terminating in a tricritical
point for general coupling in the BCS-BEC crossover. A secondary
result of our work is the demonstration that the NSR scheme yields
unphysical results in a broad region around unitarity. This is
significant, as it is widely viewed as offering a smooth, albeit
uncontrolled approximation throughout the crossover. We emphasize
that there is no \emph{a priori} reason to believe in the accuracy
of the NSR scheme without introducing an additional parameter, as we
have done here. The Ginzburg criterion governing the smallness of
fluctuation corrections is satisfied in both the BCS limit where it
takes the form $(T_c/\varepsilon_F)^2\ll 1$, and in the BEC limit
where $k_F a\ll 1$ is the relevant criterion. But at unitarity the
shift in the transition temperature relative to the mean field value
will be of order $\varepsilon_F$. At the same time the upper
critical dimension at the tricritical point is three, so we may
expect that our results there will be little changed.

Finally, we have argued that these tricritical points play an
important role in experiments on trapped Fermi gases (see, also, the
subsequent related work on trapped gases at unitarity by Gubbels et
al.~\cite{gubbels2006}). Indeed, a recent comprehensive study of the
temperature dependence of the phase-separated state at unitarity has
yielded experimental results consistent with the phase diagram
outlined here~\cite{partridge2006_2}.


\begin{acknowledgments}
We are grateful to P. B. Littlewood for stimulating discussions, and
J. Keeling for help with the numerics.  This work has been supported
by EPSRC.
\end{acknowledgments}



\end{document}